\documentclass{ws-ijmpe}

\begin{document}

\markboth{Gao Chan Yong et al.}{t-$^{3}$He flow}

%%%%%%%%%%%%%%%%%%%%% Publisher's Area please ignore %%%%%%%%%%%%%%%
\catchline{}{}{}{}{}
%%%%%%%%%%%%%%%%%%%%%%%%%%%%%%%%%%%%%%%%%%%%%%%%%%%%%%%%%%%%%%%%%%%%

\title{TRITON-$^{3}$HE RELATIVE AND DIFFERENTIAL FLOWS AND THE HIGH DENSITY BEHAVIOR OF NUCLEAR SYMMETRY ENERGY}

\author{GAO-CHAN YONG}

\address{Institute of Modern Physics, Chinese Academy of
Sciences,  Lanzhou 730000, P.R. China\\
yonggaochan@impcas.ac.cn}

\author{BAO-AN LI}

\address{Department of Physics and Astronomy, Texas A\&M University-Commerce, Commerce, TX
75429-3011, USA\\
Bao-An\_Li@tamu-commerce.edu}

\author{LIE-WEN CHEN}

\address{Department of Physics, Shanghai Jiao Tong University, Shanghai
200240, China\\
Center of Theoretical Nuclear Physics, National Laboratory of
Heavy-Ion Accelerator, Lanzhou, 730000, China\\
lwchen@sjtu.edu.cn}

\maketitle

\begin{history}
\received{(received date)}
\revised{(revised date)}
%\accepted{(Day Month Year)}
%\comby{(xxxxxxxxxx)}
\end{history}

\begin{abstract}
Using a transport model coupled with a phase-space coalescence
after-burner we study the triton-$^3$He relative and differential
transverse flows in semi-central $^{132}Sn+^{124}Sn$ reactions at
a beam energy of $400$ MeV/nucleon. We find that the triton-$^3$He
pairs carry interesting information about the density dependence
of the nuclear symmetry energy. The t-$^3$He relative flow can be
used as a particularly powerful probe of the high-density behavior
of the nuclear symmetry energy.
\end{abstract}

\section{Introduction}

The density dependence of nuclear symmetry energy especially at
supra-saturation densities is among the most uncertain properties
of neutron-rich nuclear matter\cite{Kut94,Kub99}. However, it is
very important for nuclear structure\cite{Bro00,Hor01}, heavy-ion
reactions\cite{LiBA98,LiBA01b,Dan02a,Bar05,CKLY07,LCK08} and many
phenomena/processes in astrophysics and cosmology
\cite{Sum94,Lat04,Ste05a}. Heavy-ion reactions especially those
induced by radioactive beams provide a unique opportunity to
constrain the symmetry energy at supra-saturation densities in
terrestrial laboratories. Various probes using heavy-ion reactions
have been proposed in the literature, see, e.g., ref.\cite{LCK08}
for the most recent review. It is particularly interesting to
mention that, besides many significant results about the symmetry
energy at sub-saturation densities, see, e.g.,
refs.\cite{chen04,li05,chen05,She07,Tsang09,Cen09,Leh09},
circumstantial evidence for a rather soft symmetry energy at
supra-saturation densities has been reported very
recently\cite{xiao09} based on the IBUU04 transport
model\cite{IBUU04} analysis of the $\pi^-/\pi^+$ data taken by the
FOPI Collaboration at SIS/GSI\cite{Rei07}. To constrain tightly
and reliably the nuclear symmetry energy especially at
supra-saturation densities, much more efforts by both the nuclear
physics and the astrophysics communities are still needed.

There is an urgent need to verify the conclusion about the soft
symmetry energy at supra-saturation densities required to
reproduce the FOPI $\pi^-/\pi^+$ data within transport model
analyses\cite{xiao09,Rei07}. It was predicted that the
neutron-proton differential flow is another sensitive probe of the
high-density behavior of the nuclear symmetry energy\cite{ba00a}.
However, it is difficult to measure observables involving
neutrons. One question often asked by some experimentalists is
whether the triton-$^3$He pair may carry the same information as
the neutron-proton one. We will try to answer this question
quantitatively by coupling the IBUU04 calculations to a
phase-space coalescence after-burner. Indeed, we found that,
similar to the neutron-proton pair, the triton-$^{3}$He relative
and differential transverse flows are sensitive to the
high-density behavior of the nuclear symmetry energy. They can be
used to test indications about the high-density behavior of the
symmetry energy observed earlier from analyzing the $\pi^-/\pi^+$
data.

\section{The theoretical models}
Our study is carried out based on the IBUU04 version of an isospin
and momentum dependent transport model and the simplest
phase-space coalescence after-burner. The single nucleon potential
is one of the most important inputs to BUU-like transport models
for nuclear reactions. In the IBUU04 transport model, we use a
single nucleon potential derived within the Hartree-Fock approach
using a modified Gogny effective interaction (MDI)
\cite{IBUU04,das03}. The corresponding MDI symmetry energy can be
found in Ref. \cite{Xu09}

Because most BUU-type transport models including the IBUU04 are
incapable of forming dynamically realistic nuclear fragments, some
types of after-burners, such as statistical and coalescence
models, are normally used as a remedy. This kind of hybrid models
can be used to study reasonably well, for instance, nuclear
multifragmentation, see, e.g.,
ref.\cite{kruse85,Ligross,hagel,Tan01}, collective flow of light
fragments, see e.g., \cite{koch90,chen98,zhang99} and the
formation of hypernuclei\cite{gait08}. There are, however, some
remaining issues, such as the freeze-out time of fragments that is
related to the time of coupling the transport model with the
after-burner, etc. There are also interesting work in using
advanced coalescence models\cite{Mat97,Ru99}, see, e.g.,
refs.\cite{chen03,Chen04b}. We notice here that, several advanced
cluster recognition routines, such as, the Early Cluster
Recognition Algorithm (ECRA) \cite{Str97}, the Simulated Annealing
Clusterization Algorithm (SACA) \cite{Pur00}, have been put
forward in recent years. For the purposes of the present
exploration, however, we use the simplest phase-space coalescence
model, see, e.g., refs.\cite{chen98,zhang99}, where a physical
fragment is formed as a cluster of nucleons with relative momenta
smaller than $P_{0}$ and relative distances smaller than $R_{0}$.
The results presented in the following are obtained with
$P_{0}=263$ MeV/c and $R_{0}=3$ fm. This simple choice may thus
limit the scope and importance of our study here. For instance, we
shall limit ourselves to studies of the relative/differential
observables for t-$^3$He pairs without attempting to study pairs
of the heavier mirror nuclei. An extended study including the
heavy mirror nuclei using the advanced coalescence and/or earlier
cluster recognition methods is planned.

\section{Results and discussions}

\begin{figure}[th]
\centerline{\psfig{file=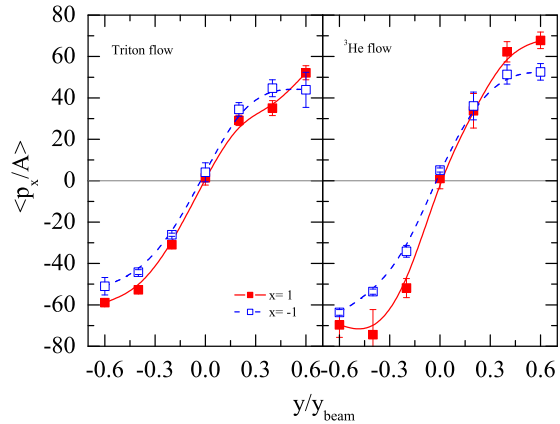,width=8cm}} \vspace*{8pt}
\caption{Triton and $^{3}$He transverse flows (in unit of MeV) as
functions of the reduced C.M. rapidity in the semi-central
$^{132}Sn+^{124}Sn$ reactions at a beam energy of $400$
MeV/nucleon. Taken from Ref. 41.} \label{resflow}
\end{figure}
We now investigate whether the transverse collective flows of
triton and $^{3}$He can be used to probe the symmetry energy.
Firstly, we examine in Fig.\ \ref{resflow} their transverse flows
individually. The average C.M. transverse momentum per nucleon
$<p_{x}/A>$  in the reaction plane is defined as
\begin{eqnarray}
<p_{x}/A>(y) &\equiv
&\frac{1}{N(y)}\sum_{i=1}^{N(y)}p_{x}^{i}/A(y)
\end{eqnarray}
where $N(y)$ is the total number of fragments of mass A in the
rapidity bin at $y$. The correlation between the $<p_{x}/A>$ and
rapidity $y$ reveals the transverse collective flow\cite{Dan85}.
It is seen that $^{3}$He clusters show a stronger flow than triton
clusters. This is mainly due to the stronger Coulomb force
experienced by the $^{3}$He clusters. More interestingly, the
transverse flow of $^{3}$He clusters show appreciable sensitivity
to the variation of the symmetry energy.

\begin{figure}[th]
\centerline{\psfig{file=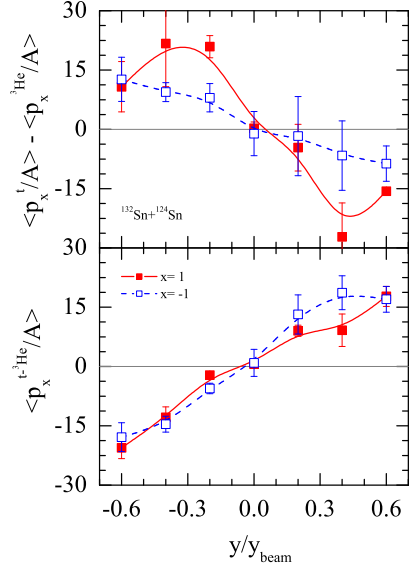,width=6cm}} \vspace*{8pt}
\caption{Triton-$^{3}$He relative and differential flows (in unit
of MeV) as a function of the reduced C.M. rapidity in the same
reaction as in Fig.\ \ref{resflow}. Taken from Ref. 41.}
\label{clusterflow}
\end{figure}

The transverse flow is a result of actions of several factors
including the isoscalar, symmetry and Coulomb potentials and
nucleon-nucleon scatterings. It is well known that the transverse
flow is sensitive to the isoscalar potential. Given the remaining
uncertainties associated with the isoscalar potential and the
small size of the symmetry energy effects, it would be very
difficult to extract any reliable information about the symmetry
energy from the individual flows of triton and $^{3}$He clusters.
Thus techniques of reducing effects of the isoscalar potential
while enhancing effects of the isovector potential are very
helpful \cite{ba00a,Yong67,Li06,gre03,fami06}. We thus study in
Fig.\ \ref{clusterflow} the triton-$^{3}$He relative and
differential flows. The relative flow is given as
\begin{eqnarray}
<p_{x}^{t}/A>-<p_{x}^{^{3}\mathrm{He}}/A>=
\frac{1}{N_{t}}\sum_{i=1}^{N_{t}}p_{x}^{i}/A-
\frac{1}{N_{^{3}\mathrm{He}}}\sum_{i=1}^{N_{^{3}\mathrm{He}}}p_{x}^{i}/A.
\end{eqnarray}
The triton-$^{3}$He differential flow  reads
\begin{eqnarray}
&&<p_{x}^{t-^{3}\mathrm{He}}/A>=
\frac{1}{N_{t}+N_{^{3}\mathrm{He}}}(\sum_{i=1}^{N_{t}}p_{x}^{i}/A-\sum_{i=1}^{N_{^{3}\mathrm{He}}}p_{x}^{i}/A)\nonumber\\
&=&\frac{N_{t}}{N_{t}+N_{^{3}\mathrm{He}}}<p_{x}^{t}/A>
-\frac{N_{^{3}\mathrm{He}}}{N_{t}+N_{^{3}\mathrm{He}}}<p_{x}^{^{3}\mathrm{He}}/A>,
\label{diflow}
\end{eqnarray}
where $N_{t}$, $N_{^{3}\mathrm{He}}$ are the number of triton and
$^{3}$He in the rapidity bin at $y$. From the upper panel of Fig.\
\ref{clusterflow}, it is seen that the triton-$^{3}$He relative
flow is very sensitive to the symmetry energy. Because of the
larger slope of the $^{3}$He flow, the triton-$^{3}$He relative
flow shows a negative slope at mid-rapidity. Effects of the
symmetry energy on the differential flow shown in the lower panel,
however, is relatively small. Although the $^{3}$He flow is more
sensitive to the symmetry energy, the small number of $^{3}$He
clusters makes the $^{3}$He flow contributes less to the
triton-$^{3}$He differential flow (as indicated in Eq.
(\ref{diflow})). The triton-$^{3}$He differential flow is
therefore dominated by triton clusters. Consequently, it is less
sensitive to the symmetry energy than the triton-$^{3}$He relative
flow. The slope $F(x)\equiv d<p_x/A>/d(y/y_{beam})$ of the
transverse flow at mid-rapidity can be used to characterize more
quantitative the symmetry energy effects. We found that for the
t-$^3$He relative flow, $F(x=1)\approx-74$ MeV/c and
$F(x=-1)\approx-22$ MeV/c, respectively. For the t-$^3$He
differential flow, $F(x=1)\approx21$ MeV/c and $F(x=-1)\approx42$
MeV/c, respectively. The t-$^3$He relative flow thus can be used
as a very useful and independent tool to test the soft symmetry
energy at supra-saturation densities extracted from studying the
$\pi^-/\pi^+$ ratio\cite{xiao09}.

\section{Summary}

In summary, using a hybrid approach coupling the transport model
IBUU04 to a phase-space coalescence after-burner we studied the
t-$^3$He relative and differential flows in semi-central
$^{132}Sn+^{124}Sn$ reactions at an incident energy of $400$
MeV/nucleon. We found that the nuclear symmetry energy affects
strongly the t-$^3$He relative and differential flows. The
t-$^3$He relative flow can be used as a particular powerful probe
of the high-density behavior of the nuclear symmetry energy.

\section{Acknowledgements}

This work was supported in part by the US National Science
Foundation Awards PHY-0652548 and PHY-0757839, the Research
Corporation under Award No.7123 and the Texas Coordinating Board
of Higher Education Award No.003565-0004-2007, the National
Natural Science Foundation of China under grants 10710172,
10575119, 10675082 and 10975097 and MOE of China under project
NCET-05-0392, Shanghai Rising-Star Program under Grant
No.06QA14024, the SRF for ROCS, SEM of China, and the National
Basic Research Program of China (973 Program) under Contract
No.2007CB815004.

\end{document}